\documentclass[12pt,english,a4paper]{article}
\usepackage{natbib}
\usepackage{rotating,placeins}
\usepackage{babel}
\usepackage{geometry}
\usepackage[latin1]{inputenc}
\usepackage[T1]{fontenc}
\usepackage{graphicx,graphics,mathpple,textcomp}
\usepackage{fancyhdr}
\usepackage{lscape,epsfig,amssymb,amsmath}
\usepackage[]{subfigure}
\usepackage{rotating,placeins}
\usepackage{lscape}
\usepackage{setspace}
\title{Nearshore wave forecasting and hindcasting by dynamical and statistical 
downscaling}
\author{Øyvind Breivik\footnote{Corresponding author. E-mail: \texttt{oyvind.breivik@ecmwf.int}}
\footnote{Published as Breivik, {\O}, Y Gusdal, B~R Furevik, O~J Aarnes and M Reistad, 2009:
Nearshore wave forecasting and hindcasting by dynamical and statistical downscaling, 
\emph{J Marine Syst}, \textbf{78}, S235--S243, doi:10.1016/j.jmarsys.2009.01.025}
\footnote{Address: Norwegian Meteorological Institute, All\'{e}g 70, NO-5007
Bergen, Norway}, Yvonne Gusdal, Birgitte R Furevik, Ole Johan Aarnes, \\
and Magnar Reistad}
\date{Available online 3 March 2009}

\begin{document}

\maketitle
\begin{abstract}
A high-resolution nested WAM/SWAN wave model suite aimed at rapidly
establishing nearshore wave forecasts as well as a climatology and return
values of the local wave conditions with Rapid Enviromental Assessment (REA)
in mind is described. The system is targeted at regions where local wave
growth and partial exposure to complex open-ocean wave conditions makes
diagnostic wave modelling difficult.

SWAN is set up on 500~m resolution and is nested in a 10~km version of WAM.
A model integration of more than one year is carried out to map the spatial
distribution of the wave field.  The model correlates well with wave buoy
observations (0.96) but overestimates the wave height somewhat (18\%,
bias 0.29~m).

To estimate wave height return values a much longer time series is required
and running SWAN for such a period is unrealistic in a REA setting. Instead
we establish a direction-dependent transfer function between an already
existing coarse open-ocean hindcast dataset and the high-resolution
nested SWAN model.  Return values are estimated using ensemble estimates
of two different extreme-value distributions based on the full 52 years
of statistically downscaled hindcast data.  We find good agreement between
downscaled wave height and wave buoy observations.  The cost of generating
the statistically downscaled hindcast time series is negligible and can be
redone for arbitrary locations within the SWAN domain, although the sectors
must be carefully chosen for each new location.

The method is found to be well suited to rapidly providing detailed wave
forecasts as well as hindcasts and return values estimates of partly sheltered
coastal regions.
\end{abstract}

\noindent
Subject keywords: rapid environmental assessment, nearshore wave forecasting,
wave hindcasting, statistical downscaling, dynamical downscaling, wave height
return values.  \\ 
Regional terms: Europe, Norway. North Sea.  \\

\section{Introduction}
Coastal regions partially sheltered from the open ocean prove a difficult
middle ground between open-ocean conditions and the really small scales found
in harbours and the mouths of rivers and estuaries.  Prognostic wave models
developed for open-ocean conditions \citep[see e.g.][]{has88,tol91} have
until recently been considered too computer-intensive to be operated on grids
resolving complex coastal regions (typically requiring a grid finer than 1~km).

In partially sheltered domains where local wave growth is limited and the ocean
spectrum outside sheltering islands can be assumed spatially homogeneous it
may be possible to use simple refraction-diffraction models \citet{ore93},
especially if the main concern is relatively uni-directional low-frequency
swell from distant storms.  In such cases it may be possible to establish
a tractable database (look-up table) of relations between open-ocean and
nearshore conditions for various combinations of integrated parameters
like significant wave height, peak direction and peak period of the wave
field impinging on the boundary of the model domain.  On even finer scale
semi-diagnostic models designed for harbours, closed bays and estuaries
\citep[e.g. STWAVE,][]{smi01b} perform very well.

However, for coastal domains where local wave growth is of significance the steady-state
assumption breaks down. Also, the complexity of the sea state near the open
ocean with swell intrusion and young wind sea requires full two-dimensional
spectra as boundary conditions to the fine-scale model. This
is difficult to achieve with steady-state models because the aforementioned
look-up table will grow out of bounds (the ``curse of dimensionality''),
making a fully non-stationary dynamical spectral wave model a computationally
competitive alternative.

The advent of high-resolution prognostic wave models specifically designed to
handle the high resolution needed to resolve nearshore conditions combined
with numerical weather prediction models (NWP) capable of capturing the
complexity of the coastal wind field opens up the possibility of forecasting
the sea state in regions partly sheltered from the open ocean on spatial
resolution of less than 1~km. \emph{Simulating WAves Nearshore} (SWAN) is a
third-generation wave model \citep{boo99,ris99} in operational use at
the Norwegian Meteorological Institute since 2006. The model is operated on
500~m resolution and is used to issue wave forecasts to the Norwegian Coastal
Administration and the general public for particularly sensitive sea areas.

For this study SWAN was set up for a region on the west coast of Norway which
is partially sheltered by islands to the north-west and with a larger island
to the east, see Figures~\ref{Fig:modelsys} and \ref{Fig:kart}.  We define a
semi-sheltered coastal region as one that is exposed to wind and waves from
the open ocean and is large enough for local wave growth to become important
while still being sheltered by islands or mainland in other directions.

Running a high-resolution forecast system with SWAN as the wave component is
computationally demanding, but tractable, in forecast mode. If detailed wind
fields are also required (depending on the steepness of the topography),
a detailed numerical weather prediction model must also be operated. In
our case very detailed (4-5~km resolution) winds were used to force the
wave model. The wind fields are taken from a high-resolution nested weather
prediction system. Full two-dimensional wave spectra from a coarser wave model
(WAM) are used as boundary conditions for the high-resolution model. This
nested setup with full two-dimensional wave spectral information on the
open boundaries will be referred to as the \emph{dynamical} downscaling.
Dynamical downscaling techniques for waves resemble the nesting methods
employed for atmospheric and oceanographic modelling but with the important
difference that the wave field is a forced dynamical system that depends
solely on the wind field, the open boundary and the bathymetry. Thus, while
a nested numerical weather prediction model or an ocean model would generate
small-scale phenomena (eddy activity) that could not be predicted from the
boundary values alone, the wave model will only respond to structures in the
fine-scale wind field and details in the bathymetry that were not resolved
by the coarser model.

With rapid environmental assessment (REA) in mind, the next step once a
forecast system is in place will be to create annual and seasonal maps of the
wave field, e.g. significant wave height and dominant wave direction. This
can to a good estimate be achieved with hindcast simulations of intermediate
length, typically one year. However, extending the high-resolution wave
model integration to generate a hindcast archive covering \emph{decades}
is computationally prohibitive on the spatial resolution described above,
at least for REA purposes.  To estimate the extremes (return values) of the
wave climate in coastal locations we build a direction-dependent statistical
transfer function between a coarse-resolution open-ocean hindcast archive
(covering the period 1955 and onwards) and the high-resolution coastal
SWAN domain for the overlapping time period.  This is referred to as the
\emph{statistical} downscaling. Local wave growth and local wind effects
(land-sea breeze, funneling along the coast) as well as sheltering from
nearby islands are processes in the coastal zone that can significantly
alter the local wave conditions compared with the wave field in the open ocean.
To account for this, our approach is to relate the offshore conditions
from the coarse hindcast archive to the sea state in the coastal location
found with SWAN for the common time period through a direction-dependent
transfer function.  A detailed study of the relative merits of statistical
and dynamical downscaling of waves to nearshore conditions can be found
in \citet{gas06} or \citet{gas06b}. A detailed development of
statistical downscaling techniques is found  in \citet{sto99}.

The objective of this study is to outline and evaluate an approach to rapid
assessment of wave conditions in coastal locations with complex topography
and complex sea state. We will assess a method for quickly setting up a
reliable forecasting system as well as building a hindcast (climatology)
series of sufficient length to properly estimate the average and the extremes
of the wave climate.  The method to be evaluated can be summarized as follows.
\begin{itemize}
\item Set up a nested high-resolution prognostic wave model capable of
precisely forecasting and recreating the wave conditions in coastal,
semi-sheltered waters where local wave growth and exposure to the complex
open ocean wave conditions are important.  Force the model with detailed wind
fields if necessary (depending on the steepness of the topography and the
complexity of the coastline).
\item Run the model for a sufficiently long period to assess the forecast skill
and to map the fine-scale spatial variations in \emph{average} wave climate,
i.e., the first and second moments of the wave field. This involves typically a one-year
integration.
\item Use this ``training'' period to build a transfer function to open-ocean
hindcast series and calculate extremes (return values) of the wave height
distribution from the transfer function for chosen locations.  
\end{itemize}

\section{Dynamical downscaling October 2005-January 2007}
\subsection{WAM50 and WAM10 open ocean wave models}
\begin{figure}[h]
\begin{center}
\includegraphics[scale=0.4]{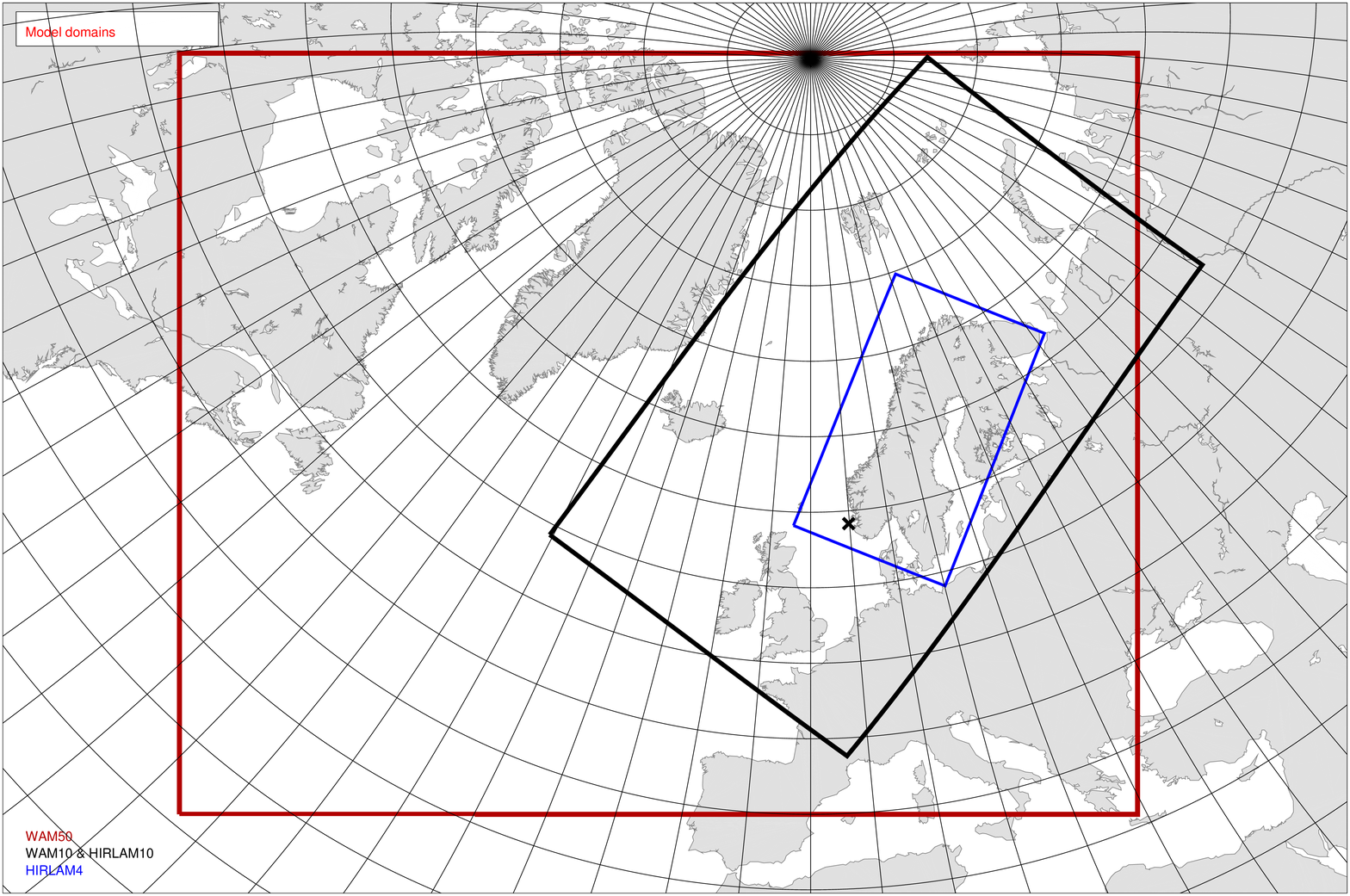}
\end{center}
\caption{\label{Fig:modelsys}
The model domains of the operational models used for providing boundary
conditions for the SWAN dynamical downscaling during the 16-month period. The
numerical weather prediction model HIRLAM4 covers Scandinavia and only the
eastern part of the Norwegian Sea (blue).  HIRLAM10 and the wave model WAM10
share the same domain and grid resolution (black) while WAM50 covers a large
part of the North Atlantic to ensure that swell is properly accounted for
(red). The location of the high-resolution SWAN model shown in Fig.~\ref{Fig:kart} is 
marked with ``x''. $5^\circ$ graticule, rotated plate carr\'{e}e projection.}
\end{figure}

The third-generation wave model WAM \citep{has88} has been in
operational use at the Norwegian Meteorological Institute since 1998.
A medium-resolution (10~km) domain (hereafter referred to as WAM10) is
nested in a coarse-resolution model covering the North Atlantic on 50~km
grid resolution (hereafter referred to as WAM50).  The model discretizes
the two-dimensional spectrum with 24 directional bins and 25 logarithmically
spaced frequency bins covering the range from 0.042 to 0.4~Hz. This covers
the energetic part of the open-ocean wave spectrum.  All the wave model
implementations described in this work have been forced with 10~m wind from
the emph{High Resolution Limited Area Model} (HIRLAM) suite of numerical
weather prediction models in operational use at the Norwegian Meteorological
Institute \citep{und02}. The model domains are nested: HIRLAM20 (20~km
resolution), HIRLAM10 (10~km resolution) and HIRLAM5 (5~km resolution).
In December 2005 HIRLAM10 and HIRLAM20 were upgraded from version 6.2 to 6.4,
while the 5~km implementation was upgraded 1 June 2006 and exchanged for a
4~km grid (hereafter referred to as HIRLAM4).  The model domains of HIRLAM10,
HIRLAM4/5, WAM50 and WAM10 are shown in Fig.~\ref{Fig:modelsys}.

The standard WAM configuration constrains nested models to fit perfectly inside
each other. This means that the coarse domain must have a grid spacing that
is divisible by the grid spacing of the fine-resolution domain, i.e., $\Delta
x_\mathrm{c}/\Delta x_\mathrm{f}$ is integer.  For the long Norwegian coastline
facing the North Sea and the Norwegian Sea this is cumbersome and represents
a major computational cost as the ideal orientation of the 10~km-resolution
model on a rotated spherical grid is along the major axis of the coastline
toward the north east (see Fig.~\ref{Fig:modelsys}).  To circumvent this
problem, the WAM nesting scheme was exchanged for a more flexible setup
where the boundary file is constructed from the spectral output (2D) of
the outer model which is interpolated and rotated to be aligned with the
boundary of the inner model.  The nesting scheme handles longitude-latitude
(plate carr\'{e}e) as well as rotated spherical grids and is used both for
nesting WAM10 in WAM50 as well as for SWAN inside WAM10.

\subsection{SWAN 500~m nearshore model}
\begin{figure}[h]
\begin{center}
  \includegraphics[scale=1.2]{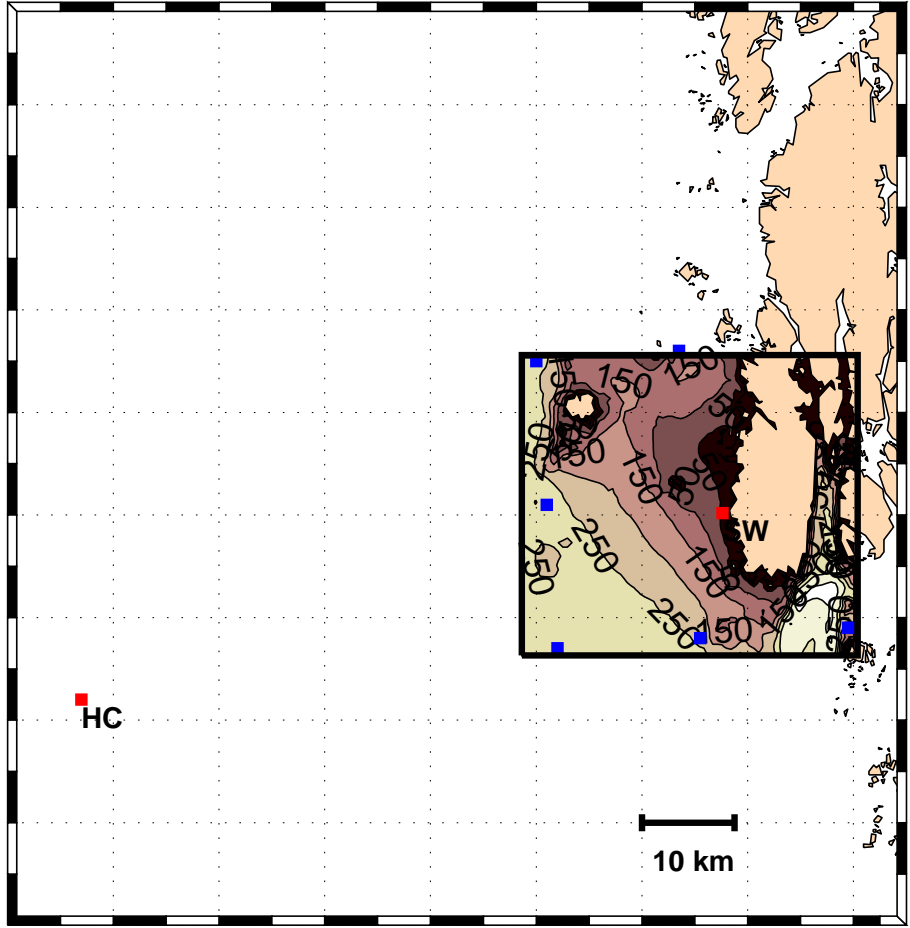}\\
  \caption{The SWAN model domain (500~m resolution, inner black box).
  The buoy location is marked as ``SW''. The hindcast grid point is marked as
  ``HC''. Blue dots indicate the location of WAM10 2D spectra used as boundary
  conditions for SWAN. The bathymetry of the SWAN domain is indicated with depth contours
  for every 50~m. The domain is only marginally affected by bottom refraction and friction.}
  \label{Fig:kart}
  \end{center}
\end{figure}

SWAN is a third generation prognostic spectral wave model which includes
shallow water effects, such as depth-induced wave breaking, friction and
triad wave-wave interaction \citep{boo99,ris99}. It is thus well suited
for detailed coastal modelling.  Alternatively, the shallow-water version
of WAM could have been nested into the deep-water open-ocean model suite
\citep{mon00}, but as our high-resolution domain is quite deep (50-300~m
depth, see Fig.~\ref{Fig:kart}) wave damping is negligible.  On the other
hand, the short time step required to comply with the Courant-Friedrichs-Lewy
criterion on such small spatial scales becomes important, and the main reason
for selecting SWAN is its fully implicit numerical scheme which allows longer
time steps than the semi-implicit scheme used by WAM \citep{boo99}.

The model is set up on a 500~m resolution grid using 10~m winds from
HIRLAM5 and later HIRLAM4, both nested into HIRLAM10.  The spectrum is
discretized in 36 directional bins and 33 logarithmically spaced frequency
bands spanning 0.046-1~Hz.  The model is nested in WAM10 and interpolates 2D
spectra to all open boundary points from the WAM10 grid points indicated in
Fig.~\ref{Fig:kart}. A spectral interpolation from the 24 directional bins
and 25 frequency bins of WAM10 is performed during model integration.

SWAN was integrated over a 16-month period from 1 October 2005 to 31
January 2007.  Spectra from a new WAM hindcast archive on 10~km resolution
for the domain indicated in Fig.~\ref{Fig:modelsys} from 1957 and onwards
\citep[see][]{rei07} were used as boundary conditions, thus reducing
the computational effort to a simple integration of SWAN.

\begin{figure}[h]
\begin{center}
  \includegraphics[scale=0.86]{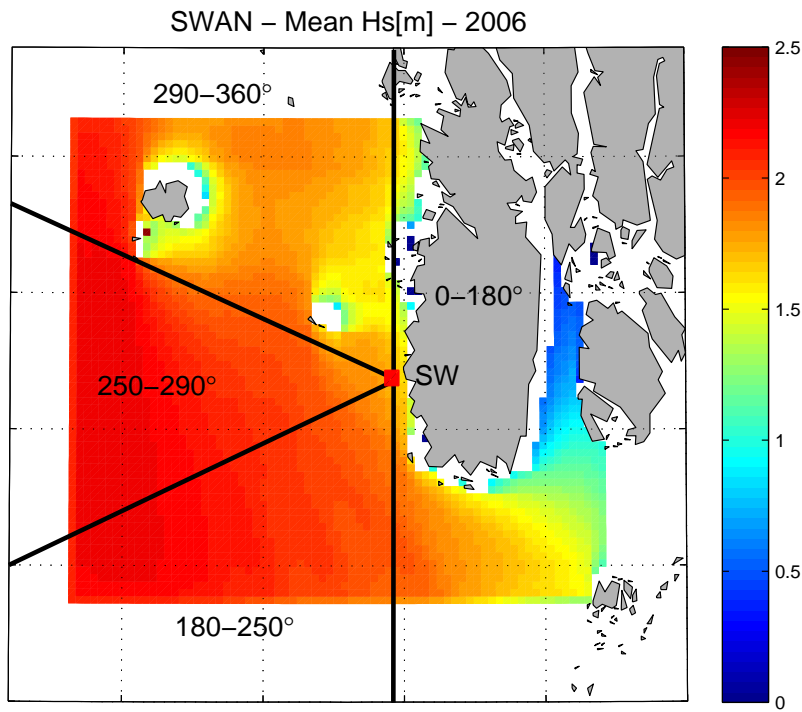}\\
(a)\\
  \includegraphics[scale=0.97]{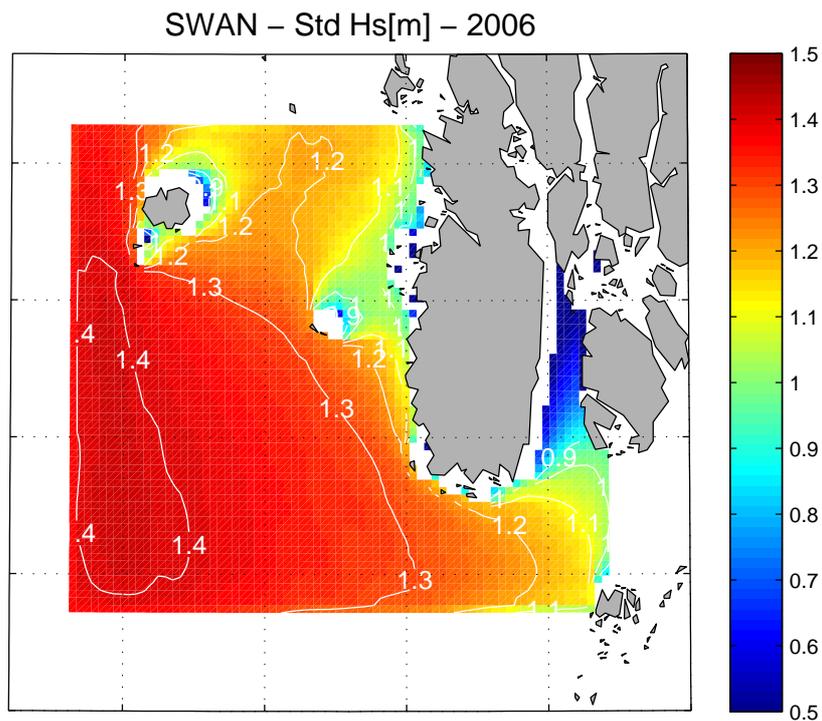}\\
(b)\\
  \caption{Panel (a): Directional binning of the nearshore location. Four
  sectors have been chosen to account for (1) the complete sheltering from the
  east, (2) the open-ocean conditions with waves propagating at a steep angle
  from the south and south-west, (3) the completely unsheltered conditions from the west
  and south-west and (4) the partial sheltering from the north and north-west
  by islands. The peak wave direction in the hindcast grid point is used
  as the selection criterion. The SWAN annual average $\overline{H_\mathrm{s}}$
  for 2006 is seen to vary significantly throughout the model domain. The buoy
 location is marked as ``SW''. This position is used for the statistical
 downscaling of the offshore hindcast data. Panel (b): Standard deviation
 of annual $H_\mathrm{s}$.}
  \label{Fig:sectors} 
\end{center}
\end{figure}

Fig.~\ref{Fig:sectors} shows the annual mean and the standard deviation
of the significant wave height for the year 2006.  The annual mean,
$\overline{H_\mathrm{s}}$, is approximately 2.5~m on the western boundary
of the model domain. This is reduced to 1.7~m at the buoy location
(marked ``SW''). It is important to note that this is not due to bottom
refraction and dissipation, as the water depth lies in the range 50-250~m
(see Fig.~\ref{Fig:kart}). Rather, the attenuation of the mean significant
wave height is the result of the partial sheltering from the north-west and
the south-east.

\begin{figure}[h]
\begin{center}
 \includegraphics[scale=0.9]{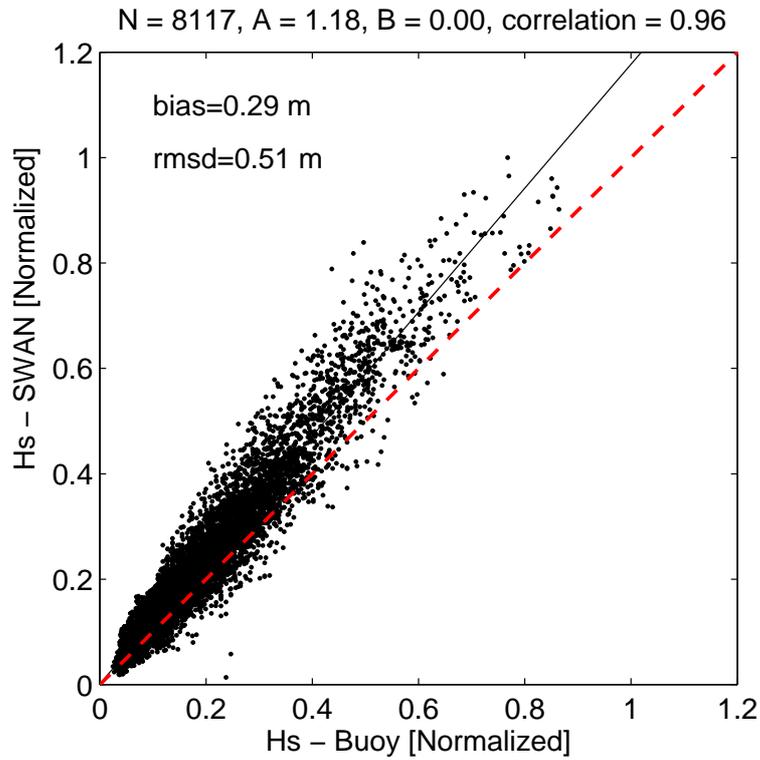}\\(a)\\
 \includegraphics[scale=0.9]{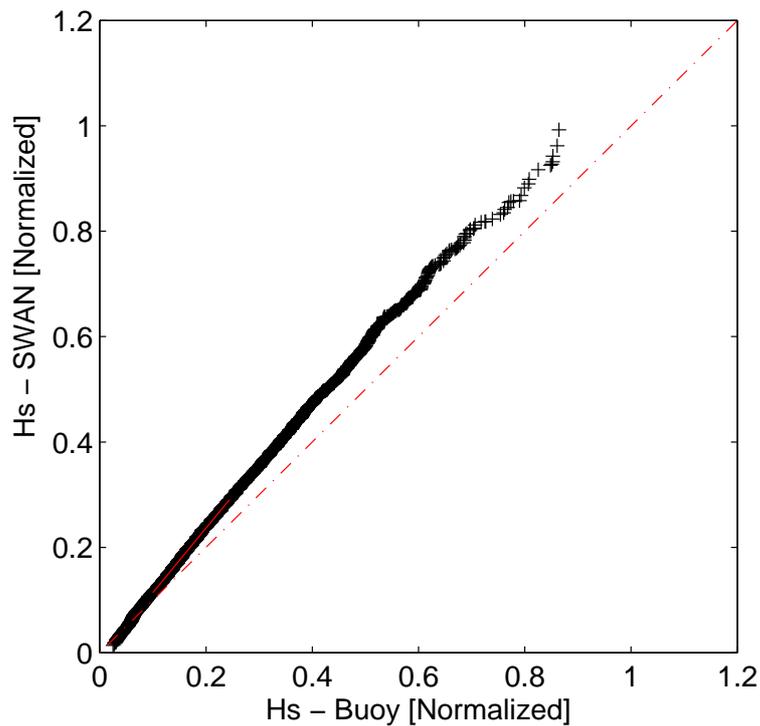}\\(b)
  \caption{Normalized scatter plot and quantile-quantile plot of wave buoy
  $H_\mathrm{s}$ vs SWAN in location ``SW''. Data were collected in two periods
  with two different buoys. The first buoy operated between 1 October and 12
  November 2005. A new buoy was installed on 25 March 2006 and was in operation
  until 23 January 2007. The correlation found between model and wave
  buoy is high (0.96), but SWAN overestimates the wave height on average by 18\%
  (bias 0.29~m).}
  \label{Fig:SWANvsBUOY}
\end{center}
\end{figure}

Buoy measurements from the period 1 October to 12 November 2005 and 25 March
2006 to 23 January 2007 in location marked ``SW'' (see Fig.~\ref{Fig:sectors})
have been compared with SWAN. The winter months from November 2005 to March
2006 are missing.  The buoy experienced some technical difficulties during the
second period and some data filtering has been carried out to remove erroneous
data.  The model agrees on average well with the buoy (correlation $r=0.96$),
but SWAN wave heights exceed the buoy measurements (Fig.~\ref{Fig:SWANvsBUOY})
by 18\% (mean bias 0.29~m).  SWAN also underestimates the wave period $T_{m02}$,
with a bias of -0.35~s (correlation $r=0.79$).  Whether the wave height bias
is due to under-estimation by the buoy or over-estimation by the model is
not clear, but waves coming from the exposed sector ($250-290^\circ$) are
virtually unbiased compared with the WINCH hindcast archive in the offshore
location (see Panel (c), Fig.~\ref{Fig:HCSvsSWAN} and discussion below),
suggesting that the wave buoy may be underestimating the true wave height
somewhat. We stress that the measurement series is short and is meant for
independent evaluation of the model setup.

\section{Statistical downscaling of hindcast series 1955-2006}
\subsection{WINCH hindcast archive}
The Norwegian Meteorological Institute maintains a coarse-resolution hindcast
data set covering the period 1955 and onwards. The hindcast archive is
generated with a second generation wave model \citep[WINCH, see][]{gre85}
with winds calculated in part from digitized pressure maps.  The model was set
up on a coarse (150~km to 75~km) grid which covered the northern North Atlantic,
the Norwegian Sea, the Greenland Sea, the Barents Sea and the North Sea. The
archive originally covered the period 1955-1981, but has since been updated
regularly \citep{rei98}.  Although by now surpassed by more sophisticated
models and forcing fields \citep[see][]{rei07}, the need for statistical
stationarity in error terms still makes it worthwhile to maintain and update
the archive. Here we employ the time series from the nearest grid point to
perform a statistical downscaling to nearshore conditions.

\subsection{Sector-wise linear downscaling}
\begin{figure}[h]
\begin{center}
\begin{tabular}{ccc}
  (a)\includegraphics[scale=0.55]{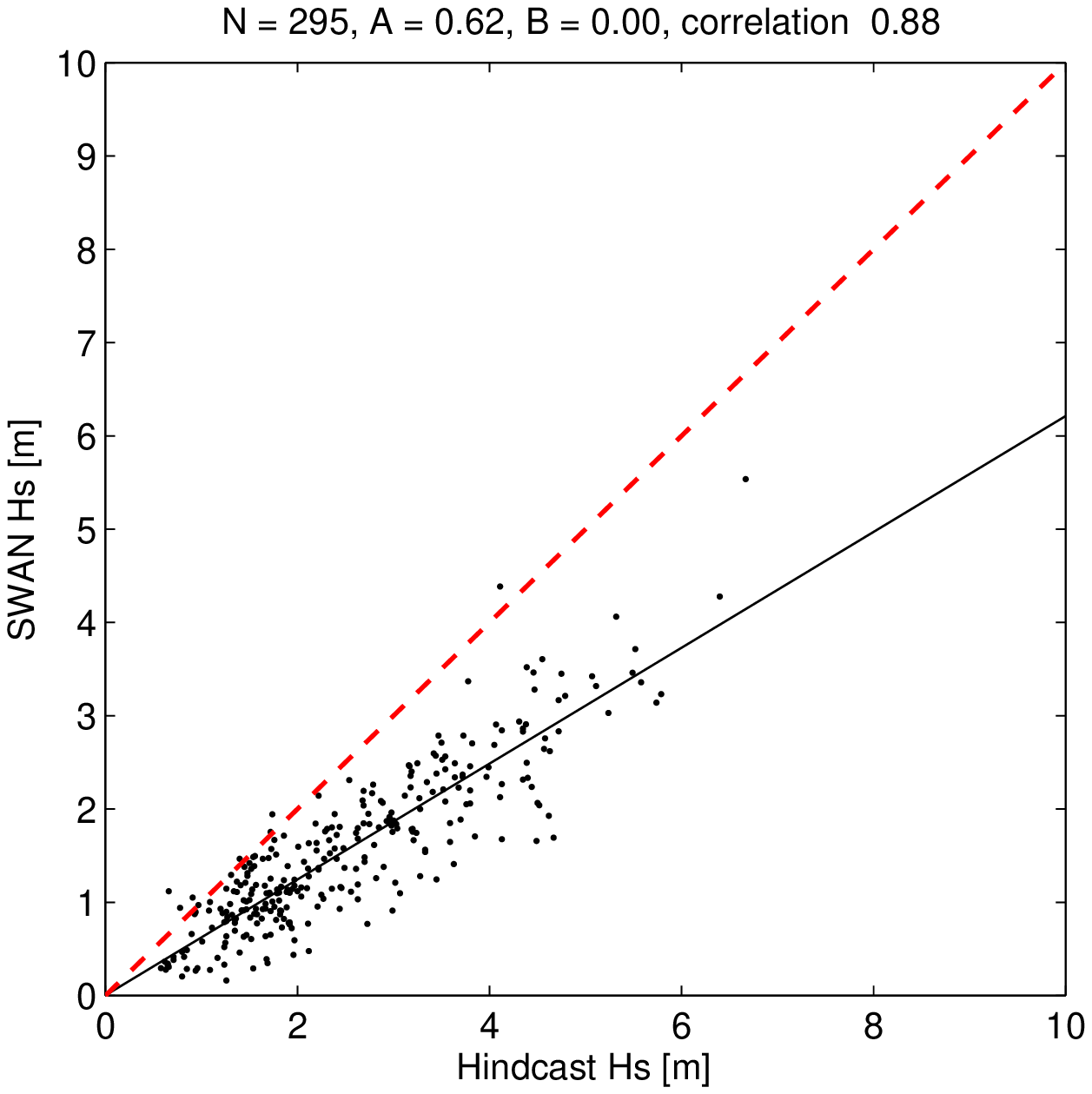}&
  (b)\includegraphics[scale=0.55]{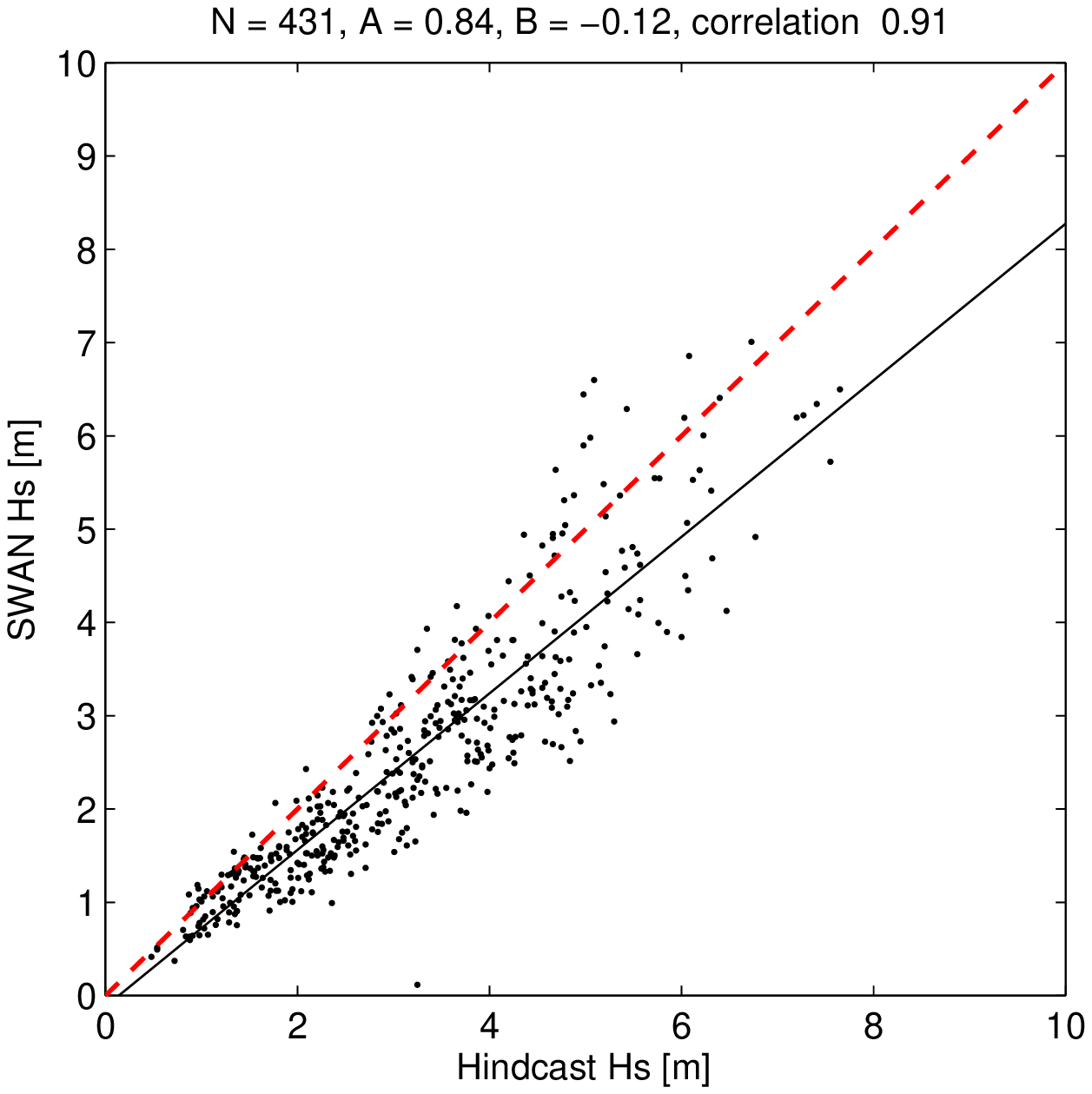}\\
\\
  (c)\includegraphics[scale=0.55]{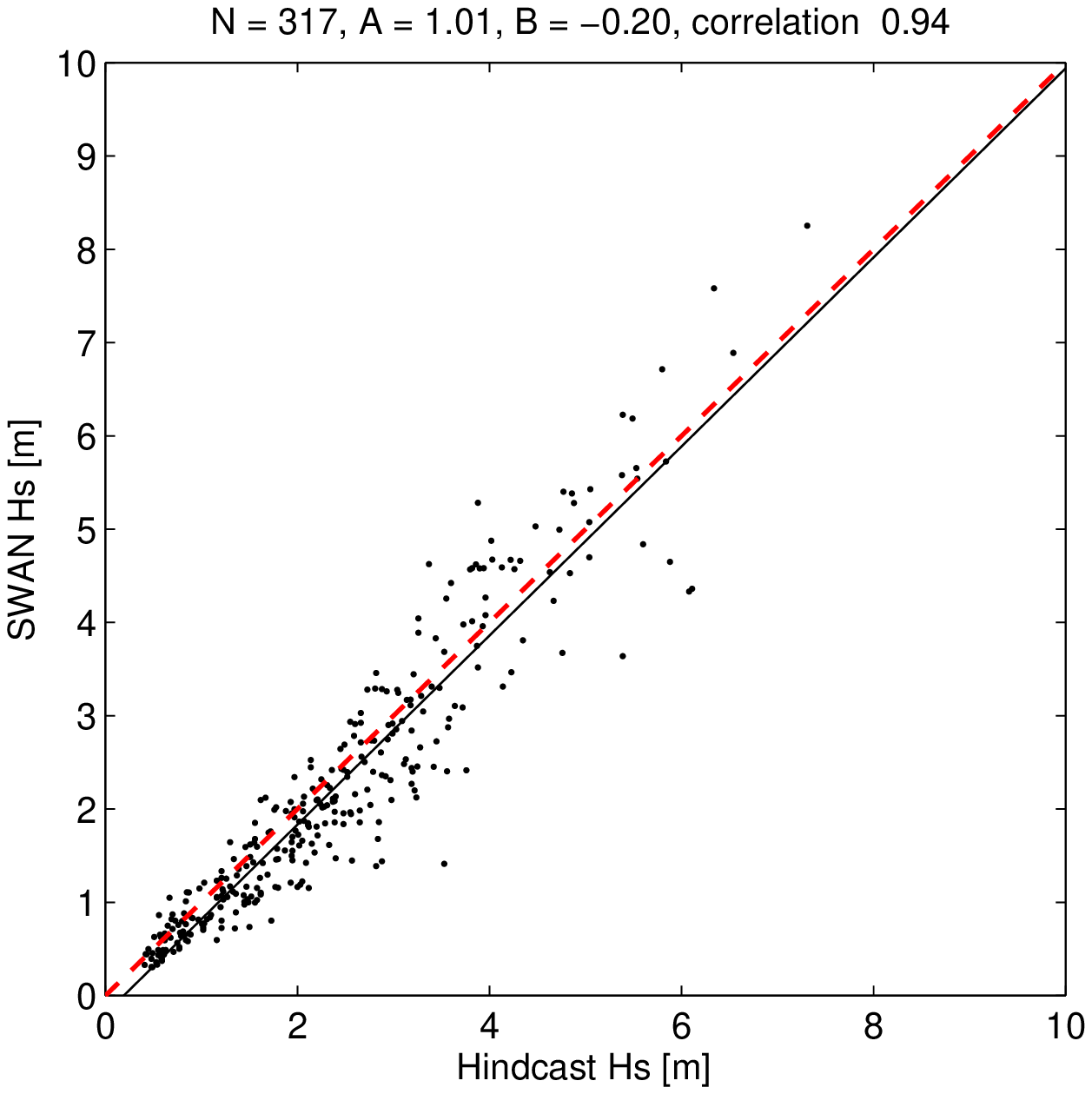}&
  (d)\includegraphics[scale=0.55]{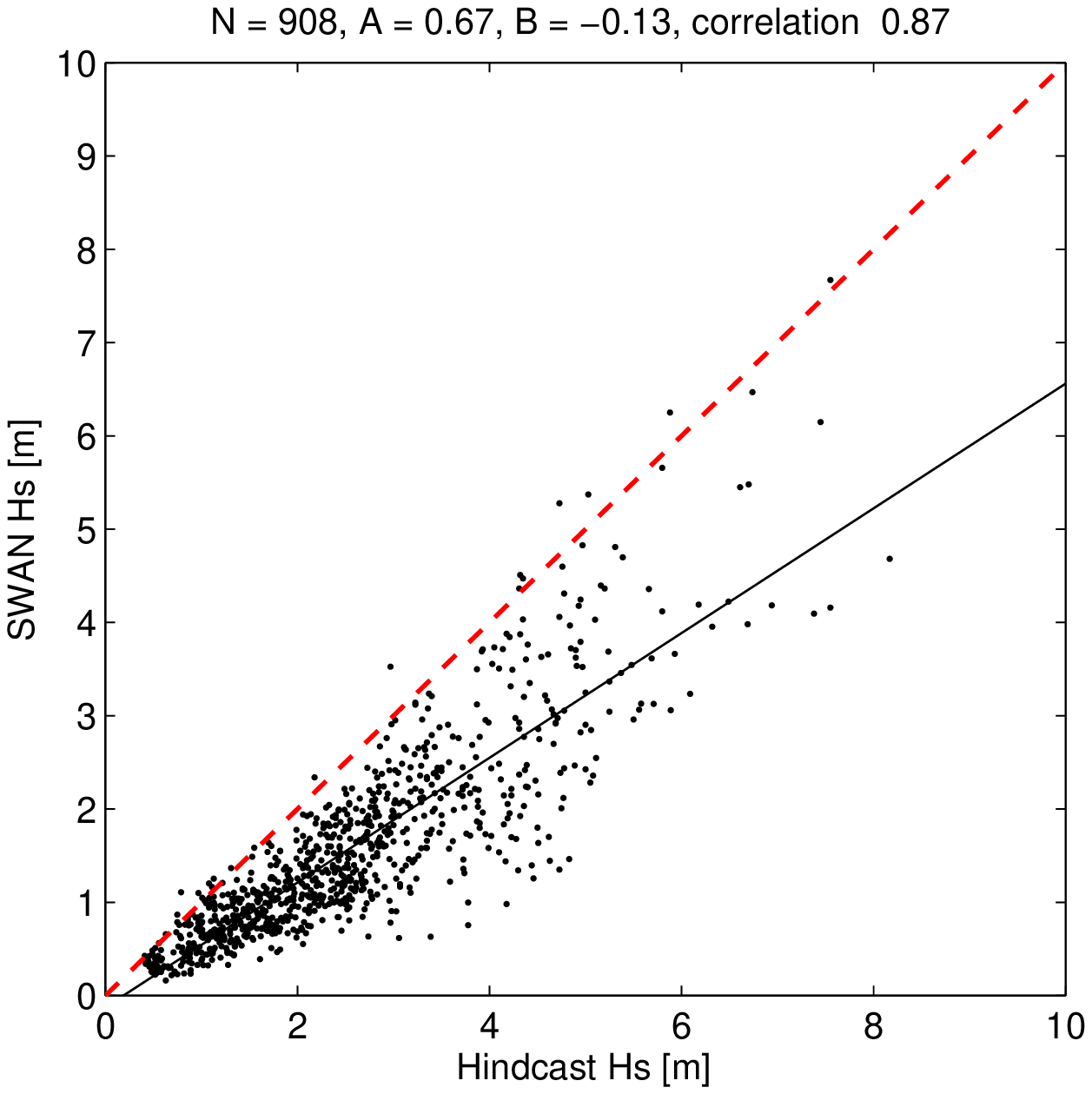}\\
\\
  (e)\includegraphics[scale=0.55]{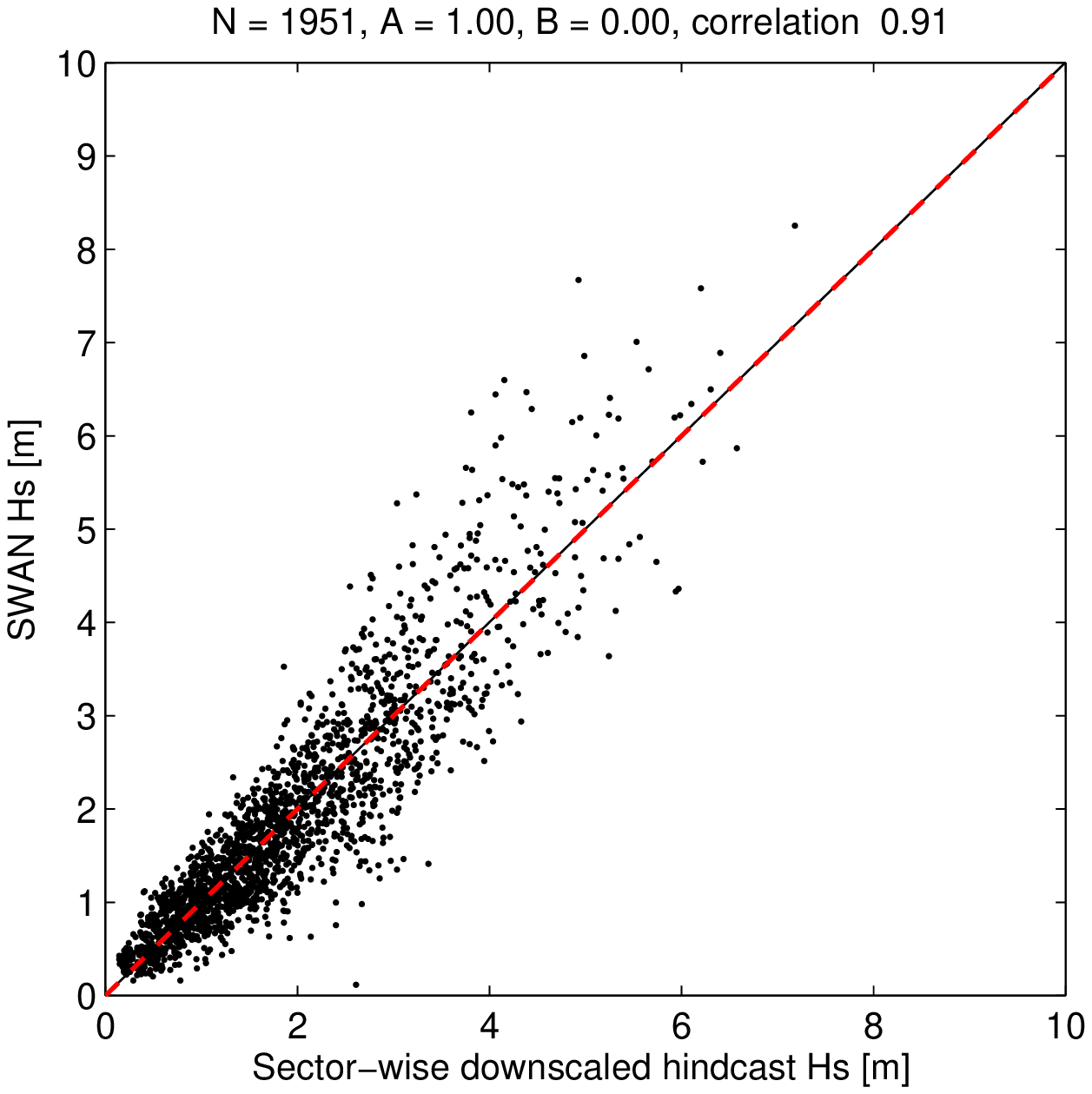}&
  (f)\includegraphics[scale=0.55]{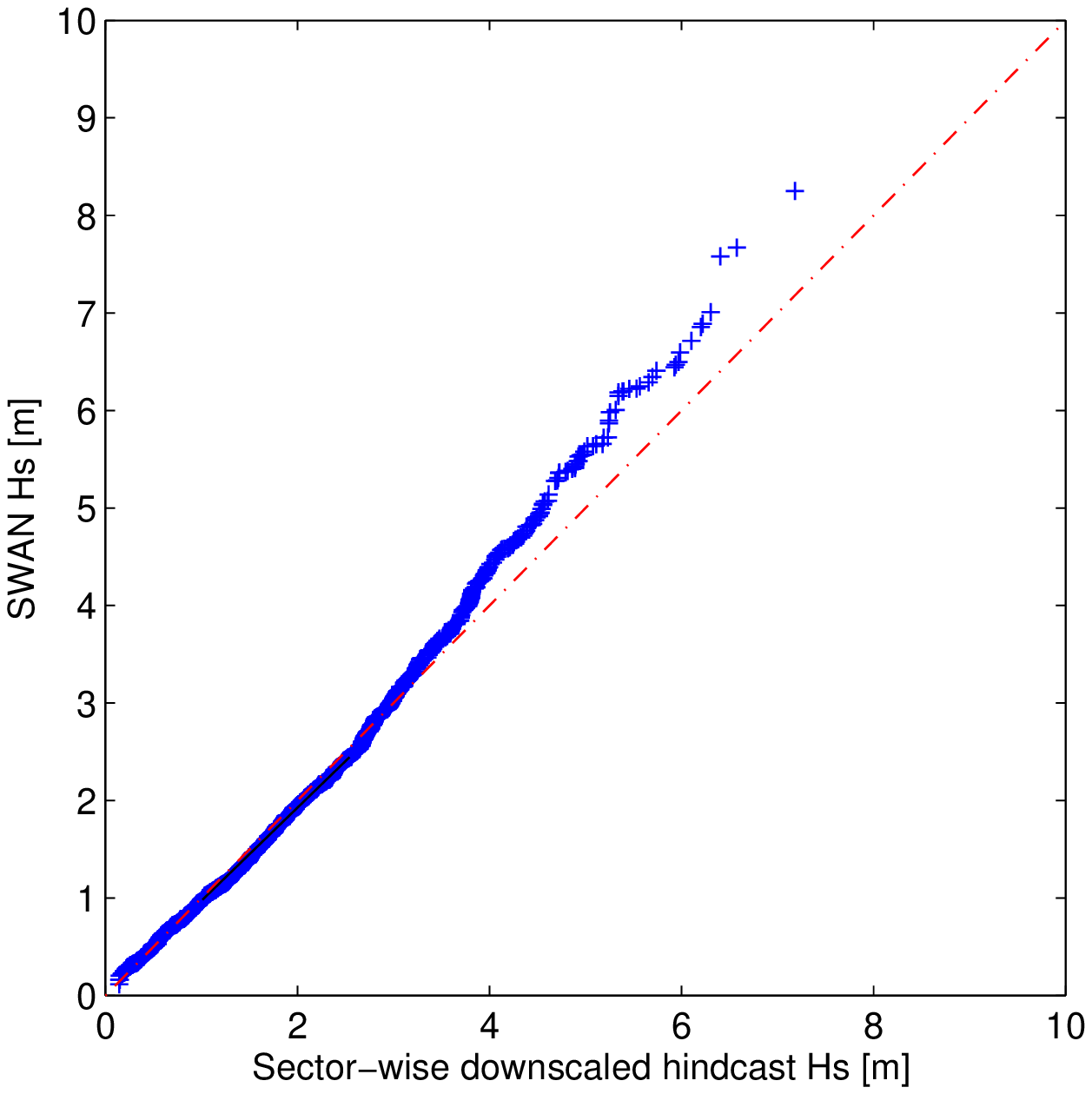}\\
\end{tabular}
\caption{Sector-wise downscaling of hindcast $H_{\mathrm{s}}$ from the
open-ocean location (HCS) to the SWAN nearshore location (coincident with buoy
location). Panels (a)-(d) show good agreement with SWAN for the four
sectors (representing $0-180^\circ$, $180-250^\circ$, $250-290^\circ$ and
$290-360^\circ$, respectively), with correlation coefficients ranging from
0.88 to 0.94.  Panel (e) shows the overall agreement (combined correlation
0.91) between the sectorwise downscaling of the hindcast data and the
SWAN nearshore location. The rms spread is used to add Gaussian noise when
estimating the return values.  The quantile-quantile plot for the combined
downscaling from all four sectors is shown in Panel (f), revealing a slight
under-estimation of the highest waves compared with SWAN.}
\label{Fig:HCSvsSWAN} 
\end{center} 
\end{figure}

To account for the partial sheltering by islands  a direction-dependent
linear regression has been performed to relate the significant wave height
found in the open-ocean hindcast location to the SWAN wave height found in
the nearshore buoy location  (see Fig.~\ref{Fig:sectors}). The directional
binning is based on the peak wave direction ($\theta_\mathrm{p}$) in the
hindcast location. The wind direction could also be used as the binning
criterion, but the wind can change rapidly and because of inertia in the wave
field, wave and wind directions may in some situations differ significantly.
Four directional bins have been chosen (Fig.~\ref{Fig:sectors}), based on
considerations of the differential sheltering for the various directions
toward the nearshore location (coincident with buoy location).  For each
sector (bin) $j$, the linear regression can be written as
\begin{equation}
y_{j}=A_{j}x_{j}+B_{j}+e_{j},
\label{eq:linreg}
\end{equation}
where $j$ represents sectors 1, 2, 3, and 4. A time series of downscaled
wave height $y_{j}$ is related to the offshore hindcast wave height,
$x_{j}$, through relation (\ref{eq:linreg}) for each sector $j$. An
aggregated time series $h$ of nearshore wave height is since constructed
from these four transfer functions. The rms error $s$ of the aggregated
transfer function is used to add Gaussian noise $e_{j} \in N(0,s)$ to
the downscaled time series \citep{wil95}.  The transfer functions are
shown in Panels (a)-(d) in Fig.~\ref{Fig:HCSvsSWAN} (summarized in Table
\ref{tab:transferfunction}). The impact of sheltering on the nearshore wave
conditions and how the directional binning modifies the relation between
offshore and nearshore wave height is evident. The largest reduction of
wave height relative to offshore conditions appears not surprisingly in
sectors $0-180^{\circ}$ and $290-360^{\circ}$. The wave height is here on
average reduced to 62$\%$ and 67$\%$ of the offshore hindcast wave height,
respectively. Least reduction is found in the sector $250-290^{\circ}$
(Fig.~\ref{Fig:HCSvsSWAN}), where the wave height distribution nearshore is
nearly identical to the offshore hindcast distribution.  The total (aggregated)
sector-wise transfer function is referred to as HCS.  The downscaling shows
good agreement with SWAN with an overall correlation coefficient $r=0.91$
(Panel (e) of Fig.~\ref{Fig:HCSvsSWAN}).

\begin{figure}[h]
\begin{center}
  \includegraphics[scale=0.8]{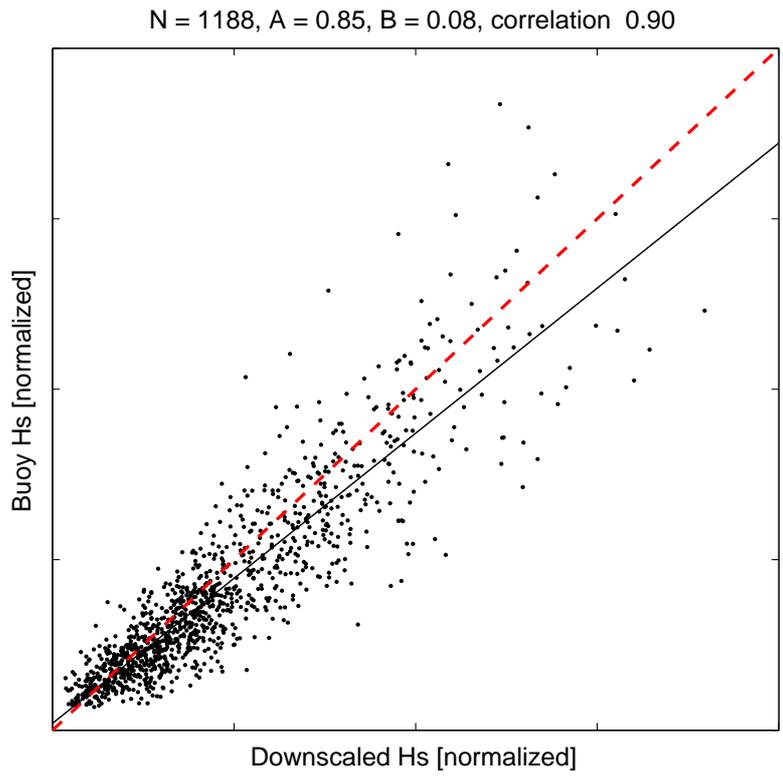}\\
(a)\\
  \includegraphics[scale=0.8]{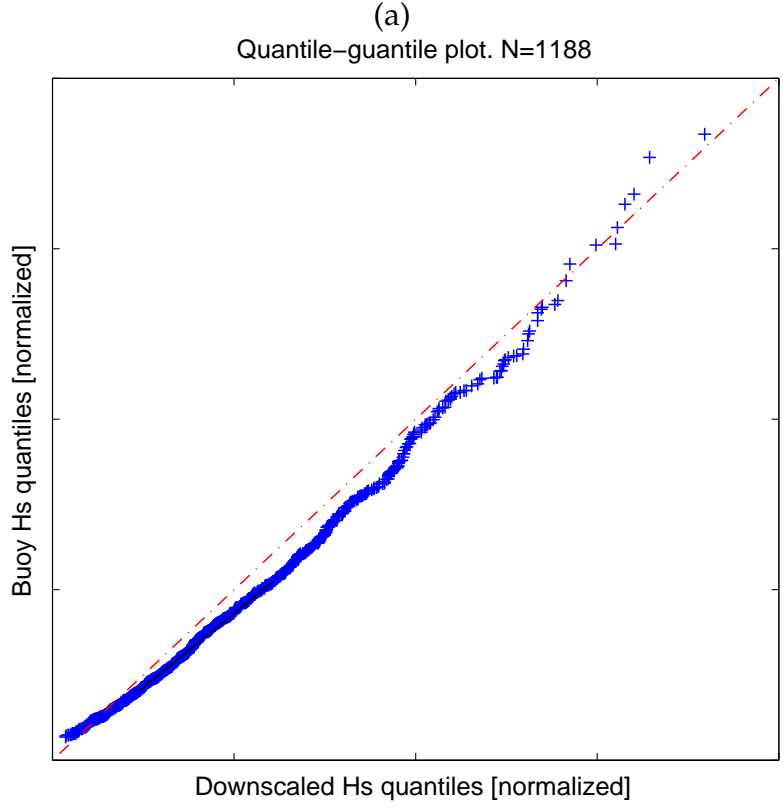}\\
(b)\\
  \caption{A comparison between the sector-wise 
  downscaling (HCS) and buoy measurements of $H_{\mathrm{s}}$. Panel (a)
  shows generally good agreement with a correlation coefficient of 0.90,
  although the downscaling tends to over-estimate the wave height slightly. 
  Panel (b): quantile-quantile plot.} 
  \label{Fig:NHCvsBUOY2}
\end{center} \end{figure}

The buoy data are used for independent evaluation of the statistical
downscaling from hindcast location to the nearshore conditions at the buoy
location.  As seen in Panel (a) of Fig.~\ref{Fig:NHCvsBUOY2}, the correlation
is high and comparable to the correlation against SWAN ($r=0.90$), although
again a slight over-representation of the wave height is found (see also
Fig.~\ref{Fig:SWANvsBUOY}). This is to be expected as the downscaling relates
the hindcast data to the SWAN data, hence any bias in SWAN is inherited
by the downscaling.  From the quantile-quantile plot in Panel (b) a slight
under-representation of the very highest waves is found. Otherwise the match
is good.

\subsection{Return values of significant wave height}\label{Sec:returverdi}
The most commonly used method for estimating return values is the
``peaks-over-threshold'' (POT) technique \citep{smi90b}, where only maximum
values exceeding a certain threshold ($u$) are kept. The assumption that the
individual data points represent maxima from individual storms is usually
ensured by requiring a minimum distance of 48 hours between consecutive data
points \citep{sim96}.
Let $\lambda$ represent the average number of storms per year exceeding the
POT threshold $u$. Then the probability of not exceeding the $R$-year return
value ($H_{R}$) in a storm is
\begin{equation}
F(H_{R})\equiv \mathrm{Prob}(h \leq H_R)=1-\frac{1}{\lambda \ R}\,.
\end{equation}
Thus, the probability of not exceeding for example the 100-year return value ($H_{100}$) is
$$
F(H_{100})=1-\frac{1}{100 \lambda}.
$$
The return value $H_R$ should be interpreted as the value that will \emph{on
average} be exceeded only once in $R$ years. \citet{wil74} notes that for
the Gumbel distribution, the probability of actually seeing a 100-year return
value in a random 100-year period is approximately 63\%. Similar values apply
for the other extreme-value distributions discussed below. The threshold $u$ is manually
chosen and should be high relative to typical values. \citet{nae01}
suggest that optimal results are obtained if the threshold is chosen
so that the number of exceedances is approximately 10 per year, while
\citet{lop00} suggest a threshold 2-3 times the mean value of
$H_\mathrm{s}$. Here maxima from storms exceeding a threshold of 5~m are kept
(5-7 storms per year).

Assuming that the wave heights extracted using the POT method are independent
(representing maxima from different storms) and identically distributed
(iid), the generalized extreme value \citep[GEV,][]{jen55} distribution can be applied to
the extremes, \begin{equation} \label{eq:gev} \mbox{GEV}(h;a,b,c)=\left\{
\begin{array}{ll}
                             \exp\left[-(1+c\frac{h-b}{a})^{-1/c}_{+}\right]
                             & \mbox{if $c\neq 0$} \\ \\
                             \exp\left[-e^{-(h-b)/a}\right] & \mbox{if $c=0$}
                            \end{array}
         \right.
\end{equation} 
Here $h$ is the significant wave height, $a>0$ and $b$ are scale and location
parameters while $c$ determines the shape of the GEV distribution. The cases
$c>0$, $c=0$ and $c<0$ correspond to the Fr\'{e}chet (Fisher-Tippett Type II),
Gumbel (Fisher-Tippett Type I) and the reverse Weibull (Fisher-Tippett Type
III) distributions, respectively.  We will here estimate return values for
significant wave height using the Gumbel distribution ($c=0$).  In addition,
the three-parameter Weibull distribution (referred to as Weibull3, not to be
confused with the above-mentioned reverse Weibull or Fisher-Tippett Type III)
will be used,
\begin{equation} 
\label{eq:weibull} 
F_{w}(h)=\left\{
\begin{array}{ll}
                    1-\exp[-(\frac{h-b}{a})^{\gamma}]  & \mbox{if $h\geq b$}
                    \\ 0 & \mbox{if $h < b$}.
                  \end{array}
         \right.
\end{equation} 
For the Gumbel distribution, $a$ and $b$ are computed from the mean and
the standard deviation of the significant wave height. For Weibull3, the
location parameter $b$ is set to 4.995~m after performing an iterative best
fit procedure while $a$ and $\gamma$ are determined by linear regression.

\subsubsection{Ensemble estimates of return values}
\begin{figure}[h]
\begin{center}
\hspace{5mm}
\includegraphics[width=11.5cm]{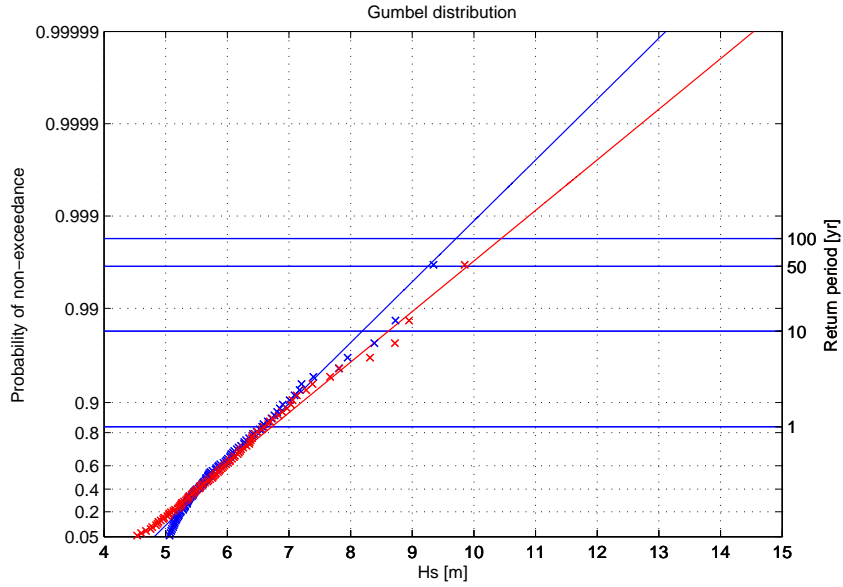}\\ (a)\\
\includegraphics[width=10cm]{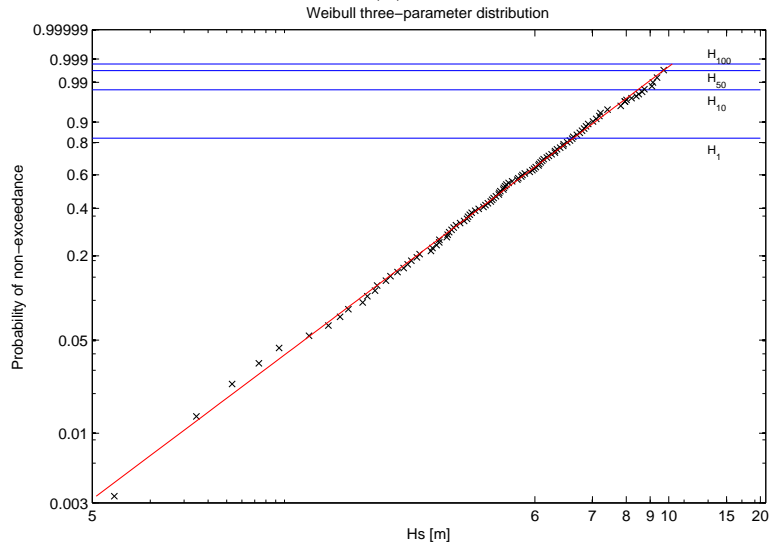}\\ (b)\\
\caption{Peaks-over-threshold estimates of the 1, 10, 50 and 100-year return
values [m] based on sectorwise statistical downscaling to nearshore conditions
of 52 years of hindcast data in an open-ocean location.  Panel (a): Gumbel
distribution. Blue curve: no noise added to downscaling (HCS).  Red curve:
Ensemble estimates with Gaussian noise consistent with rms error of the
downscaling is added (HCS$_\mathrm{e}$).  Panel (b): Three-parameter Weibull
distribution, location parameter $b=4.995$m. Blue curve: no noise. Red curve:
ensemble estimates with Gaussian noise added. Note that return values estimated
with the Gumbel distribution are more sensitive to noise than Weibull estimates,
adding 0.7~m to $H_{100}$.}\label{GUM_WEIBULL}
\end{center}
\end{figure}

Return values are
sensitive to the actual shape of the cumulative distribution as well as
individual extreme values.  To avoid under-estimating the return values
we have added Gaussian noise consistent with the rms error found in the
sector-wise statistical downscaling from the hindcast location to the SWAN
location. Return values denoted HCS$_\mathrm{e}$ in Table~\ref{tab:return}
are averages estimated from $\mathcal{O}(100)$ downscaled time series
with Gaussian random noise consistent with Eq.~(\ref{eq:linreg}) is added.
This allows us to make estimates of both the mean and the spread (standard
deviation) around the return values \citep[see also][]{nae02} for
a similar method for estimating confidence intervals on return values).
As can be seen from Table~\ref{tab:return} and from Fig.~\ref{GUM_WEIBULL}
the impact of adding noise is most pronounced for estimates using the Gumbel
distribution, where the 100-year return value goes up from 9.7~m to 10.4~m when
noise is added to the time series.  (Note that noise is added \emph{after}
extracting the peaks for the Gumbel method while noise is added \emph{before}
the selection for the Weibull3 method, since the location parameter $b$
in Eq.~(\ref{eq:weibull}) is dependent on the threshold.)  The Weibull3
method with noise gives a 100-year return value of 10.3~m and thus seems less
affected by noise (10.2~m without noise, see Table \ref{tab:return}). However,
the standard deviation around this estimate is higher than for the Gumbel
distribution.  The 100-year return value for the significant wave height
offshore is approximately 13~m (not shown). It is clear that even though the
location shown in Fig.~\ref{Fig:kart} is only partially sheltered by islands
to the north-west and the mainland to the east, the return values are lowered
significantly compared with open-ocean conditions. This reduction is not
the result of damping, keep in mind that waves from the west and south-west
arrive virtually unattenuated (Fig.~\ref{Fig:HCSvsSWAN}), but stems instead
from the fact that high waves become less likely near the coast as they can
only appear from the sector $250-290^\circ$ whereas in the open ocean high
waves can come from all directions (although the prevailing weather pattern
still makes storms from the west more likely than from the east).

\section{Assessment of method and concluding remarks}
SWAN is seen to perform well on the spatial scales of interest, i.e.,
coastal semi-sheltered conditions with a spatial grid resolution of
500~m.  The performance of the model was evaluated over two periods where
buoy measurements were available (correlation 0.96) but with a bias of
0.29~m (18\%).  Running SWAN in forecast mode is thus a good alternative
when detailed forecasts of wave conditions are required in regions where
diagnostic, steady-state models prove inadequate, provided that a coarser,
open ocean wave model is already in place.  The cost of running the model
is moderate, with two-day forecasts completed in less than one hour on a
standard Linux workstation.

We have assessed how a nested setup with WAM and SWAN can be used to map in
detail the spatial distribution of \emph{average} wave climatology for partly
sheltered coastal regions.  To do so SWAN was integrated over a 16-month
period with boundary values in the form of 2D spectra from an archived WAM
model integration.

This exercise is of course much more expensive than running short (typically
48~h) daily forecasts. It is however still tractable and the results provide
very detailed spatial maps of average sea state conditions. Such simulations
can be carried out on a standard Linux workstation in less than one week
provided wind fields and spectral boundary conditions from a coarser open-ocean wave
model system have been archived. If the full nested system need be rerun, the
cost goes up considerably. In our case, archived 2D spectra from pertinent
locations of the nested WAM50/WAM10 forecast system were available. We
emphasize that such a short (one year) model integration can only give
reliable information about the average (first moment) and the second moment
of the full sea state probability distribution (and even those moments can be
somewhat distorted if an anomalous year is sampled). To assess the tail of the
distribution, i.e., return values, a much longer integration is needed.

To estimate the extremes of the wave height distribution, the full 16-month
SWAN integration was used to build a direction-dependent transfer function
to the coarse hindcast location which in turn was used to statistically
scale down the full 52-year time series from the hindcast archive.
This allowed us a long enough series to make estimates of the extremes
of the wave climatology. Using two different extreme-value distributions,
the 100-year return value for the nearshore location was estimated to be
around 10.3~m. This is substantially lower than what is found at the open-ocean
hindcast location where the 100-year return value was found to be approximately
13~m. The cost of this additional statistical downscaling is very small, 
provided that a hindcast archive is in place and that a sufficiently long
SWAN integration has been carried out, which in our case was done anyway to
build detailed maps of average sea state conditions in the whole SWAN model
domain (Fig.~\ref{Fig:sectors}). The statistical downscaling was carried out
for one particular location (where buoy data were available for reference,
see Figures~\ref{Fig:kart} and~\ref{Fig:sectors}) using a sector-wise transfer
function for the wave height. This procedure can be repeated for any location
within the SWAN model grid at extremely little extra cost, and
deliberations over which directional bins to use represent the majority of
the work required to build the transfer function.  It is thus a very cheap
alternative to setting up a detailed diagnostic model for a region where
a prognostic high-resolution model is required to estimate spatial maps of
average wave climatology as described above. The correlation against
wave buoy measurements is naturally somewhat lower (0.90) than those found for
the SWAN simulation (0.96). More important for making realistic estimates
of nearshore return values of significant wave height is the addition
of Gaussian noise consistent with the rms error of the transfer function
(Panel (e) of Fig.~\ref{Fig:HCSvsSWAN} and Table~\ref{tab:transferfunction}).
We find that adding noise will affect the 100-year return values $H_{100}$
by as much as 0.7~m (5\%).

Dynamical downscaling of hindcast data for a semi-sheltered area using nested
prognostic models is found to yield good agreement with observations and
to be numerically tractable.  It is an efficient method to rapidly build
a detailed forecast system and to determine the average wave climatology
in locations where steady-state (diagnostic) wave models will not perform
well because of local wave growth or where they prove impractical due to the
exposure to open-ocean conditions.

The statistical downscaling for estimation of return values is numerically
very efficient and yielded good agreement with observations.  In conclusion
we found that the combination of a fully prognostic wave forecast system
which is rapidly deployable within a coarser wave forecast system and a
statistical downscaling of hindcast data based on a longer model integration
is well suited for building rapid environmental assessment systems for wave
forecasting and hindcasting in coastal regions.

\section*{Acknowledgment}
This work has been supported by the Research Council of Norway and the
Norwegian technology company FOBOX through project no 174104, ``Development
of a generic model/set of tools for prediction of waves in areas close the
coast - to be used for wave energy development''. All wave buoy measurements
have been normalized and all geographical references have been removed to
protect the intellectual property rights of FOBOX.

We also wish to thank the two anonymous reviewers for thoughtful comments
that helped us significantly to improve the manuscript.
\bibliography{/home/rd/diob/Doc/TeX/Bibtex/BreivikAbb,/home/rd/diob/Doc/TeX/Bibtex/Breivik} \newpage

\section*{Figures and tables}

\begin{table}[h]
\begin{center}
\caption{{Transfer functions between wave height offshore and nearshore
for each  sector; $s$ is the standard deviation of the Gaussian noise added.}}
\label{tab:transferfunction}
\vspace{3mm}
\begin{tabular}{|l|l|l|} \hline \hline
\textbf{Sector} & \textbf{Transfer function} & \textbf{rms error} $s$ [m]   \\[1.0ex]
\hline
$0-180^{\circ}$:       & $y = 0.62x$         & 0.40    \\[1.0ex]
$180-250^{\circ}$:     & $y = 0.84x - 0.12$  & 0.58    \\[1.0ex]
$250-290^{\circ}$:     & $y = 1.01x - 0.20$  & 0.49    \\[1.0ex]
$290-360^{\circ}$:     & $y = 0.67x - 0.13$  & 0.51    \\[1.0ex]
\hline\hline
\end{tabular}
\end{center}
\end{table}

\begin{table}[h]
\begin{center}
\caption{Peaks-over-threshold estimates of the 1, 10, 50 and 100-year
return values [m] based on sectorwise statistical downscaling to nearshore
conditions of 52 years of hindcast data in an open-ocean location.
Gumbel refers to the Gumbel distribution while Weibull3 refers to the
three-parameter Weibull distribution with location parameter $b=4.995$m.
HCS is a sectorwise downscaling with no noise added while HCS$_\mathrm{e}$
are ensemble averages of $\mathcal{O}(100)$ realizations where Gaussian noise
consistent with the rms error in the statistical downscaling was added. The
standard deviation of these estimates are included, indicating a slightly
higher spread around the Weibull3 estimate.
\label{tab:return}}
\vspace{3mm}
\begin{tabular}{|l|c c c c c|} \hline \hline
\textbf{Method} & \textbf{$H_{1}$} & \textbf{$H_{10}$} &  \textbf{$H_{50}$} & \textbf{$H_{100}$}  & Std dev $H_{100}$ [m]  \\[1.0ex]
\hline
\textbf{}
Gumbel HCS:    & 6.6 & 8.2  &  9.3  & 9.7 & N/A    \\[1.0ex]
Gumbel HCS$_\mathrm{e}$:  & 6.8  & 8.6 &  9.9 & 10.4 & 0.12   \\[1.0ex]
Weibull3 HCS:   & 6.6 & 8.5 &  9.7 & 10.2 & N/A   \\[1.0ex]
Weibull3 HCS$_\mathrm{e}$: & 6.7 & 8.4 & 9.5 & 10.3 & 0.37 \\[1.0ex]
\hline\hline
\end{tabular}
\end{center}
\end{table}
\end{document}